\documentclass[12pt,a4paper]{article}
\usepackage{epsfig}
\begin{document}
\pagestyle{empty}
\markboth{ }{ }
\newcommand{\klgl}{\:\hbox to -0.2pt{\lower2.5pt\hbox{$\sim$}\hss}
{\raise3pt\hbox{$<$}}\:}
\newcommand{\Ud}{\mbox{$U_3$}}
\newcommand{\rd}{\mbox{$\rho_3$}}
\newcommand{\rdk}{\mbox{$\rho_3(k)$}}
\newcommand{\rdr}{\mbox{$\rho_{3_R}$}}
\newcommand{\rdrk}{\mbox{$\rho_{3_R}(k)$}}
\newcommand{\ld}{\mbox{$\bar{\lambda}_3$}}
\newcommand{\ldr}{\mbox{$\bar{\lambda}_R$}}
\newcommand{\gdz}{\mbox{$\bar{g}_3^2$}}
\newcommand{\gdzr}{\mbox{$\bar{g}_3^2$}}
\newcommand{\ldk}{\mbox{$\bar{\lambda}_3(k)$}}
\newcommand{\ldrk}{\mbox{$\bar{\lambda}_R(k)$}}
\newcommand{\gdzk}{\mbox{$\bar{g}_3^2(k)$}}
\newcommand{\gdzrk}{\mbox{$\bar{g}_3^2(k)$}}
\newcommand{\ldkt}{\mbox{$\bar{\lambda}_3(k_T)$}}
\newcommand{\ldrkt}{\mbox{$\bar{\lambda}_R(k_T)$}}
\newcommand{\gdzkt}{\mbox{$\bar{g}_3^2(k_T)$}}
\newcommand{\gdzrkt}{\mbox{$\bar{g}_3^2(k_T)$}}
\newcommand{\be}{\begin{eqnarray}}
\newcommand{\ee}{\end{eqnarray}}
\newcommand{\zphi}{\mbox{$Z_\varphi(k)$}}
\markboth{ }{ }
\def\mlabel{\label}
\def\mbibitem{\bibitem}
\newsavebox{\tempbox}
\def\mycaption#1{
\sbox{\tempbox}{#1}
\vskip0.5cm
\ifdim \wd\tempbox >\hsize
#1
\else
\centering #1  
\fi
}
\def\tabcaption#1{
\par
{\centering\parbox{12cm}{\refstepcounter{table}
\mycaption{\footnotesize{\tablename\ \thetable: #1}}}
}}
\def\figcaption#1{
\par
\hspace*{15mm}{\centering\parbox{12cm}{\refstepcounter{figure}
\mycaption{\footnotesize{\figurename\ \thefigure: #1}}}
}}
\def\epsfigure#1#2{
\begin{figure}[ht]
\centering
\leavevmode
\epsffile{#1}
\figcaption{#2}
\end{figure}
}
%
\def\nepsfigure#1#2{
\begin{figure}[ht]
\centering
\leavevmode
\fbox{\parbox{1mm}{\rule{0mm}{\epsfxsize}}
\makebox[\epsfxsize]{(#1)}}
\figcaption{#2}
\end{figure}
}
\begin{flushright}
  HD-THEP-95-37
\end{flushright}
\bigskip
\begin{center}
{\huge\bf{Nonperturbative Condensates\vspace*{0.1cm}\\ }}
{\huge\bf{in the \vspace*{0.5cm}\\}}
{\huge\bf{Electroweak Phase-Transition}}
\large\bf\footnote{Talk
 given by C. Wetterich at the 3rd Colloque Cosmologie, Paris, June 7-9, 1995}
\end{center}
\bigskip
\begin{center}
 Bastian Bergerhoff\footnote{Supported by the Deutsche
 Forschungsgemeinschaft}$^,$\footnote{e-mail:
 B.Bergerhoff@thphys.uni-heidelberg.de}
 and Christof Wetterich\footnote{
 e-mail: C.Wetterich@thphys.uni-heidelberg.de} \\
 \vspace{0.5cm}
 Institut f\"ur Theoretische Physik \\
 Universit\"at Heidelberg \\
 Philosophenweg 16, D-69120 Heidelberg
\end{center}
\setcounter{footnote}{0}
\bigskip
\begin{abstract}
We discuss the electroweak phase-transition in the
early universe, using non-perturbative
flow equations for a computation of the
free energy. For a scalar mass above
$\sim 70$ GeV, high-temperature
perturbation theory cannot describe this
transition reliably. This is
due to the dominance of three-dimensional
physics at high temperatures which implies that
the effective gauge coupling grows strong
in the symmetric phase.
We give an order of magnitude-estimate of
non-per\-tur\-ba\-tive effects in reasonable
agreement with recent results from electroweak
lattice simulations.
\end{abstract}
\newpage
\noindent{\large\bf{1. High Temperature Phase-Transitions}}
\bigskip\\
There has been a lively interest in phase-transitions
in gauge-theories over the last decades since the
original work by Kirzhnits and Linde
\cite{Kir_Lin},
indicating that spontaneously broken symmetries
are restored at high temperatures.
The most
prominent examples
are the electroweak phase-transition, i.e. the restauration
of the $SU(2)_L\times U(1)_Y$-symmetry of the
standard-model, and the transition in QCD
where (approximate) chiral symmetry is restored.
These transitions are of interest for different reasons.
In the case of the electroweak phase-transition the most
prominent question is whether
the
observed baryon-asymmetry
could have been produced during such a transition
in the early universe
\cite{Delta_B}.
For this scenario to work, one needs
besides CP-violation also a
sufficiently large deviation of
baryon-number-violating processes from thermal equilibrium
\cite{Sakharov}.
This translates into the requirement that the transition
be strong enough first order.
On the other hand, if in the standard model
the transition is a second or only
weakly first order one,
or even an analytical crossover instead of a true
phase transition, an extension of the
standard model is needed. Then either
a $B-L$ asymmetry could be generated in the early universe or the
electroweak phase transition could become sufficiently
strongly first order
\cite{Delta_B}.
An answer to the question if the observed baryon-asymmetry
originated in the electroweak phase-transition
requires a detailed
understanding of the dynamics of the transition.

The interest in the chiral transition in QCD is more directly
related to speculations about the
possibility of experimental access to
the high-temperature state of QCD at heavy-ion colliders.
At present, there is no convincing evidence that such a state has
already been detected.
A better understanding of the high temperature phase would certainly
be helpful in answering the question of clean experimental
signatures
\cite{signat_of_QGP}.

For QCD there was never any question about
the need for non-perturbative methods in examining the details
of the transition.
Correspondingly there are a host of studies
with the help of lattice-simulations
or effective models.
In the case of the electroweak transition the hope that a
description by means of high-temperature perturbation theory
might be sufficient prevailed for some time.
For the most prominent qualitative features this view
seems justified for a small mass of the Higgs scalar.
For a realistic scalar mass above the experimental bound
it has been argued, however, that a quantitative
description of the high-temperature behavior and the
phase-transition is only possible by non-perturbative methods
since strong effective couplings are involved
[5-8].
This holds despite the fact that the zero temperature
electroweak interactions are weak.

The deeper reason for the breakdown of perturbation theory lies in the
effective three-dimensional character of the high-temperature
field theory
\cite{Sintra_9}.
Field theory at nonvanishing temperature $T$ can be formulated in
terms of an Euclidean functional integral where the ``time dimension''
is compactified on a torus with radius $T^{-1}$.
For phenomena
at distances larger than $T^{-1}$ the Euclidean time dimension cannot
be resolved. Integrating over modes with momenta $p^2>(2\pi T)^2$ or,
alternatively, over the higher Fourier modes on the torus (the $n\not=0$
Matsubara frequencies) leads to ``dimensional reduction'' to an effective
three-dimensional theory. This is very similar to dimensional reduction in
Kaluza-Klein theories
\cite{Sintra_11}
for gravity. The change of the effective
dimensionality for distances larger than $T^{-1}$ is manifest
in the renormalization group approach
\cite{Sintra_8}
for a computation of the temperature-dependent effective potential
or free energy.
Here one integrates over all fluctuations with momenta
$p^2>k^2$ and follows the dependence of the effective
potential on the infrared scale $k$, finally letting
$k\rightarrow0$.
The scale dependence of the effective renormalized couplings
is governed by the usual perturbative $\beta$-functions only for
$k^2>(2\pi T)^2$.
In contrast, for $k^2<(2\pi T)^2$ the running of the
couplings was found to be determined by three-dimensional
$\beta$-functions
instead of the perturbative four-dimensional ones\footnote{
Related arguments in a different
context can be found in
\cite{Sintra_12}.}.
As an alternative to integrating out all modes with $p^2>(2\pi T)^2$ an
effective three-dimensional theory for the long distance electroweak
physics
is also obtained
\cite{Sintra_13,J}
by integrating out the higher Matsubara
frequencies\footnote{For an earlier treatment of dimensional reduction
in high-temperature QCD see
\cite{Sintra_14}}.

If the three-dimensional running of the couplings becomes important, the
physics of the phase-transition is dominated by classical statistics even
in case of a quantum
field theory. A second order phase-transition is characterized
by an infinite correlation length. The critical exponents which describe
the behavior near the critical temperature are always those of the
corresponding classical statistical system. Since the fixpoints of the
three-dimensional $\beta$-functions are very different from the
four-dimensional (perturbative) fixpoints, we conclude that
high-temperature
perturbation theory is completely misleading in the vicinity of a
second order phase-transition. This argument extends to sufficiently weak
first order transitions. A second related example for the breakdown
of perturbation theory is the symmetric phase of the electroweak gauge
theory. The gauge bosons are massless in perturbation theory and the
three-dimensional running always dominates at large distances
\cite{Sintra_15}.

In order to understand the high-temperature behavior of a theory
we should understand the qualitative features of the $\beta$-functions
in three dimensions. These $\beta$-functions have nothing to do with the
ultraviolet regularization of the field theory - in this respect there is
no difference between vanishing and nonvanishing temperature. They are
rather related to the infrared behavior of the theory or the dependence
of Green functions on some sort of infrared cutoff. According to Wilson's
concept of the renormalization group these $\beta$-functions describe the
scale
dependence of the couplings if one looks at the system on larger and
larger
distances. For an understanding of systems with approximate scaling in a
certain range it is useful to define dimensionless couplings. One divides
out an appropriate power of the infrared cutoff $k$ which plays the role
of the renormalization scale. For example, the gauge
coupling $g$ in the effective three-dimensional theory is related
to the four-dimensional coupling $g_4$ and the temperature by
\be
g^2=\frac{\bar{g}_3^2}{k}=g_4^2 \frac{T}{k}
\label{Sintra1}
\ee
For the $SU(2)$-Higgs model in three dimensions relevant for the
electroweak phase transition, the dependence of
$g^2$ on the scale $k$ is given\footnote{
The coefficients depend on the precise choice of the
infrared cutoff.} by $\beta_{\bar{g}_3^2} = \frac{\partial
\bar{g}_3^2}{\partial t} = -\frac{23}{24\pi k}\bar{g}_3^4-\dots$
with $t=\ln k$
\cite{Sintra_15}. One concludes that a
non-abelian gauge theory like the electroweak theory is confining also in
three dimensions. We have depicted the running of the\vspace*{10mm}
\hspace*{0.2cm}\epsfig{file=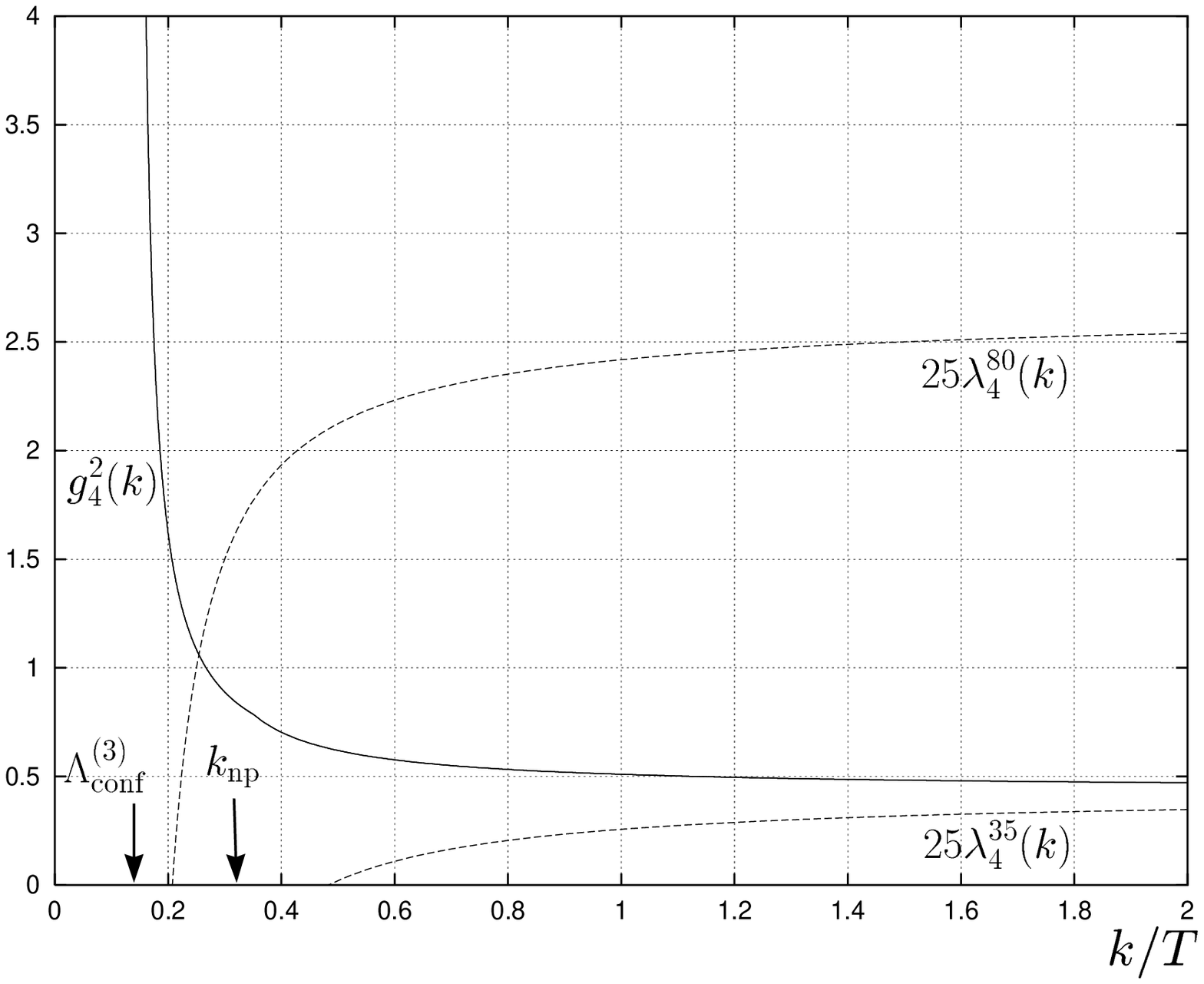,height=11cm}
\vspace*{-0.5cm}\hspace*{-0.5cm}\figcaption{The running
of the four-dimensional couplings $g_4^2$ (solid line)
and $\lambda_4$ (dashed lines). The initial values of $\lambda_4$ correspond
to scalar masses of $35$ and $80$ GeV respectively.\vspace*{1cm}}

\noindent four-dimensional
gauge coupling $g^2_4$ and
the four-dimensional quartic scalar coupling $\lambda_4$ in fig.\,1.
The deviation of $g_4^2(k)$ from the zero temperature value
$g_4^2=4/9$ can be interpreted as a measure for the validity
of the one-loop approximation.
For sufficiently small initial
$\lambda_4$
(small physical Higgs boson mass) $\lambda_4(k)$ reaches zero for $k$ much
larger
than the three-dimensional confinement scale. One then expects a
first-order
transition which is analogous to the four-dimensional Coleman-Weinberg
scenario
\cite{Sintra_16}.
Typical mass scales are of the order $k_{cw}$ where
$\lambda(k_{cw})=0$. In this case it is expected that high-temperature
perturbation theory gives reliable results. On the other hand, if the
three-dimensional confinement scale $\Lambda_{\rm conf}^{(3)}$ (the value
of $k$ for which the gauge coupling diverges or becomes very large) is
reached with $\lambda(\Lambda^{(3)}_{\rm conf})>0$ the behavior near the
phase-transition is described by a strongly interacting electroweak
theory.
Then strong effective coupling constants appear not only in the symmetric
phase, but also in the phase with spontaneous symmetry breaking.

In any case, for a calculation of interesting quantities
of a first order transition such as the critical temperature,
the nucleation rate of critical bubbles, the surface tension,
the latent heat etc. information about the strongly interacting
symmetric phase is needed. In view of this fact, there have been
different approaches to the electroweak phase-transition by means of
lattice simulations
\cite{Elweak_lattice_80,Elweak_lattice_35},
gap-equations
\cite{Elweak_gap},
effective models
and the application of renormalization group concepts
\cite{Sintra_15,ourselves}.
In the following we want to review this last approach and
present some comparison with results from $3$-dimensional
lattice simulations.
\bigskip\noindent\\
{\large\bf{2. The Average Action}}
\bigskip\\
A useful tool for describing the running of couplings in arbitrary
dimension is the average action
\cite{Sintra_18}.
Consider a simple model with a
real scalar field
$\chi$. The average scalar field is easily defined by
\be
\phi_k(x)=\int d^dyf_k(x-y)\chi(y)
\label{Sintra12}
\ee
with $f_k$ decreasing rapidly for $(x-y)^2>k^{-2}$ and properly
normalized. The average is taken over a volume of size $\sim k^{-d}$.
The average action $\Gamma_k[\varphi]$ obtains then by functional
integration of the
``microscopic variables'' $\chi$ with a constraint forcing $\phi_k(x)$ to
equal the ``macroscopic field''
$\varphi(x)$ up to small fluctuations. It is the effective action for
averages
of fields and therefore the analogue in continuous space of the block spin
action
\cite{Sintra_19}
on the lattice. All modes with momenta $q^2>k^2$ are
effectively
integrated out. Lowering $k$ permits to explore the theory at longer and
longer distances. The average action has the same symmetries as the
original action. As usual it may be expanded in derivatives, with average
potential $U_k(\rho),\rho={1 \over 2}\varphi^2$, kinetic term, etc.
\be
\Gamma_k=\int d^dx\left\{ U_k(\rho)+\frac{1}{2}
Z_k(\rho)\partial_\mu\varphi\partial^\mu\varphi+...\right\}\,\,.
\label{Sintra13}
\ee

In a suitable formulation
\cite{Sintra_20}
the effective average action corresponds to the
generating functional for 1PI Green functions with an infrared cutoff set
by
the scale $k$. It interpolates between the classical action for
$k\rightarrow\infty$ and the effective action for
$k\rightarrow0$. In this version an
exact nonperturbative
evolution equation
\cite{Sintra_20}
describes the dependence of $\Gamma_k$ on the infrared
cutoff $k$ $(t=\ln k)$
\be
\frac{\partial}{\partial t}\Gamma_k=\frac{1}{2} \int \frac{d^dq}
{(2\pi)^d}
\left(\Gamma_k^{(2)}+R_k\right)^{-1}\frac{\partial}{\partial
t}R_k\,\,.
\label{Sintra14}
\ee
Here $R_k(q)$ is a suitable infrared cutoff which may depend on $q^2$, as,
for
example,
$R_k=q^2\exp\left(-q^2/k^2\right)\left(1-\exp\left(-q^2/
k^2\right)\right)^{-1}$ or $R_k=k^2$. The
two-point function $\Gamma_k^{(2)}$ obtains by second functional variation
of $\Gamma_k$
\be
\Gamma_k^{(2)}(q',q)=\frac{\delta^2\Gamma_k}
{\delta\varphi(-q')\delta\varphi(q)}\,\,.
\label{Sintra15}
\ee
Therefore $(\Gamma_k^{(2)}+R_k)^{-1}$ is the exact propagator in presence
of
the infrared cutoff $R_k$ and the flow equation (\ref{Sintra14})
takes the form of
the scale variation of a renormalization group-improved one-loop
expression
\cite{Sintra_20_a}.
We emphasize that the evolution equation is fully
nonperturbative and no approximations are made. A simple proof can be
found
in
\cite{Sintra_20}.
The exact flow equation (\ref{Sintra14}) can be shown to be equivalent
with earlier versions of ``exact renormalization group equations''
\cite{Sintra_21}
and it encodes the same information as the Schwinger-Dyson
equations
\cite{Sintra_22}.

An exact nonperturbative evolution equation is not yet sufficient for an
investigation of nonperturbative problems like high-temperature field
theories. It is far too complicated to be solved exactly. For
practical use it is crucial to have a formulation that allows to find
reliable nonperturbative approximative solutions. Otherwise speaking,
one needs a description of $\Gamma_k$ in terms of only a few $k$-dependent
couplings. The flow equations for these couplings can then be solved
numerically or by analytical techniques. It is on the level of such
truncations of the effective average action that suitable approximations
have to be found. In this respect the formulation of the effective
average action offers important advantages: The average action has
a simple physical interpretation and eq. (\ref{Sintra14})
is close to perturbation theory if the
couplings are small. The formulation is in continuous space and all
symmetries - including chiral symmetries or gauge symmetries
\cite{Sintra_15}
 - can
be respected. Since $\Gamma_k$
has a representation as a functional integral, alternative methods
(different
from solutions of the flow equations) can be used for an estimate of its
form. Furthermore, the flow equation (\ref{Sintra14})
is directly sensitive to the relevant infrared physics
since the contribution of particles with mass larger than $k$ is
suppressed by the propagator on the r.h.s. of eq. (\ref{Sintra14}).
The closed form
of this equation does not restrict one a priori to
given expansions like in 1PI $n$-point functions.
In addition the momentum integrals in eq. (\ref{Sintra14}) are both infrared
and ultraviolet convergent if a suitable cutoff $R_k$ is chosen. Only
modes in the vicinity of $q^2=k^2$ contribute substantially. This
feature is crucial for gauge theories where the formulation of a
gauge-invariant ultraviolet cutoff is difficult without dimensional
regularization.
\newpage
\bigskip\noindent
{\large\bf{3. The Running Gauge Coupling}}
\bigskip\\
We are now ready to discuss the running of the three-dimensional
gauge coupling. We start from the effective average action for a pure
$SU(N_c)$
Yang-Mills theory. It is a gauge-invariant functional of the gauge field $A$
and obeys the exact evolution equation
\cite{Sintra_15}
(with Tr including a
momentum
integration)
\be
\frac{\partial}{\partial t}\Gamma_k[A]=\frac{1}{2} {\mbox{Tr}}
\left\{\frac{\partial R_k[A]}{\partial
t}\left(\Gamma_k^{(2)}[A]+\Gamma_k^{\rm gauge(2)}
[A]+R_k[A]\right)^{-1}\right\}-\epsilon_k[A]\,\,.
\label{Sintra16}
\ee
Here $\Gamma_k^{\rm gauge(2)}[A]$ is the contribution from a
generalized gauge-fixing term in a
covariant background gauge
and $\epsilon_k[A]$ is the ghost contribution
\cite{Sintra_15}.
The infrared cutoff
$R_k$ is in general formulated in terms of covariant derivatives. We make
the
simple truncation
\be
\Gamma_k[A]&=&\frac{1}{4}\int d^dxZ_{F,k}F_{\mu\nu}F^{\mu\nu} \nonumber \\
\Gamma_k^{\rm gauge}[A,\bar A]&=&\frac{1}{2\alpha}\int d^dxZ_{F,k}(D_\mu[\bar
A](A^\mu-
\bar A^\mu))^2
\label{Sintra18}
\ee
with background field $\bar{A}$ and $\Gamma_k^{\rm gauge(2)}[A]=
\Gamma_k^{\rm gauge(2)}[A,\bar{A}=A]$. In $d$ dimensions
the gauge coupling $\hat g$ appearing
in $F_{\mu\nu}$ and $D_\mu$ is a constant independent of $k$. The effective
$k$-dependent coupling can be associated with the dimensionless
renormalized
gauge coupling
\be
g^2(k)= k^{d-4} \bar g^2_d(k)=k^{d-4}Z^{-1}_{F,k}\hat g^2\,\,.
\label{Sintra19}
\ee
The running of $g^2$ is related to the anomalous dimension $\eta_F$
\be
\eta_F&=&-\frac{\partial}{\partial t}\ln Z_{F,k} \nonumber \\
\frac{\partial g^2}{\partial t}&=&\beta_{g^2}=(d-4)g^2+\eta_F
g^2\,\,.
\label{Sintra20}
\ee
Evaluating (\ref{Sintra16}) for configurations with
constant magnetic field and $\alpha=1$ it was found
\cite{Sintra_15}
to
obey approximately\footnote{In lowest order in the
$\epsilon$-expansion
\cite{Sintra_23}
the denominator in the last term is absent and $v_3a_3$ is replaced by
$v_4a_4$.}
\be
\frac{\partial g^2}{\partial t}=(d-4)g^2-
\frac{\frac{44}{3}N_cv_da_dg^4}
{1-\frac{20}{3} N_c v_d b_dg^2}
\label{Sintra21}
\ee
with
\be
v^{-1}_d&=&2^{d+1}\pi^\frac{d}{2}\Gamma\left(\frac{d}{2}\right) \nonumber \\
a_d&=&\frac{(26-d)(d-2)}{44}n^{d-4}_1 \nonumber \\
b_d&=&\frac{(24-d)(d-2)}{40}l^{d-2}_1\,\,.
\label{Sintra22}
\ee
Only the momentum integrals $(x\equiv q^2)$
\be
n^d_1&=&-\frac{1}{2}k^{-d}\int^\infty_0 dx\ x^\frac{d}{2} \frac{\partial}
{\partial
t}\left(\frac{\partial P}{\partial x}P^{-1}\right) \nonumber \\
l^d_1&=&-\frac{1}{2}k^{2-d}\int^\infty_0 dx\ x^{\frac{d}{2}-1}
\frac{\partial}{\partial
t} P^{-1}
\label{Sintra23}
\ee
depend on the precise form of the infrared cutoff $R_k$ appearing in
\be
P(x)=x+Z^{-1}_kR_k(x)\,\,.
\label{Sintra24}
\ee
In four dimensions one has $a_4=1,\ v_4=1/32\pi^2$ and
eq. (\ref{Sintra21}) reproduces
the one-loop result for $\beta_{g^2}$ in lowest order  $g^4$. For
the choice of $R_k$ of
\cite{Sintra_15}
where $b_4=1$ an expansion of
eq. (\ref{Sintra21}) in powers of $g^2$
also gives 93 \% of the perturbative two-loop coefficient.
We observe that the approximations leading to (\ref{Sintra21})
are valid only for $|\eta_F|<1$. For larger values of $|\eta_F|$ we
use a rougher estimate where $b_d$ is set to zero.

Concerning the high-temperature field theory we should use the
three-dimensional $\beta$-function for $k<k_T$, where
$k_T=2\pi T$ is the scale where the three-dimensional running
sets in. The ``initial
value'' of the
gauge coupling reads $g^2(k_T)=2\alpha_w(k_T)$ with $\alpha_w \approx
\frac{1}{30}$ the
four-dimensional weak fine structure constant. For $k<k_T$ the
three-dimensional
gauge coupling increases with a power behavior instead of the
four-dimensional
logarithmic behavior. The three-dimensional confinement scale
$\Lambda_{\rm conf}
^{(3)}$ - where $g^2$ diverges - is proportional to the temperature.
Similarly, we may define (somewhat arbitrarily) the scale
$k_{\rm np}$ where nonperturbative effects become
important by $|\eta_F(k_{\rm np})|=1$.
For
the
electroweak theory and the choices (\ref{Sintra24})
$P(x)=x+k^2$ ($P(x)=x/(1-\exp-
\frac{x}{k^2})$) one finds
\cite{Sintra_15}
\be
\Lambda^{(3)}_{\rm conf}&=&0.14 T \,\,\,\,(0.12 T)\nonumber \\
k_{\rm np}&=& 0.35 T \,\,\,\,(0.31 T) \,\, .
\label{Sintra26}
\ee

For the symmetric phase of the electroweak theory
one therefore has to deal with a strongly interacting gauge
theory with typical nonperturbative mass scales only somewhat below the
temperature scale! Similar to QCD one expects that condensates like
$<\!\!F_{ij}F^{ij}\!\!>$ play an important role
[6-8].
More generally, the physics of the symmetric phase  corresponds to a
strongly
coupled $SU(2)$ Yang-Mills theory in three dimensions: The relevant
excitations are ``$W$-balls'' (similar to glue balls
or strongly interacting $W$-bosons) and scalar bound
states. All ``particles'' are massive (except the ``photon'') and
the relevant mass scale is set by $\Lambda_{\rm conf}^{(3)}\sim T$.
For the $W$-boson masses we expect typical masses
between $k_{\rm np}$ and $\Lambda^{(3)}_{\rm conf}$
(cf.\,(\ref{Sintra26})). Also the
values of all condensates are given by appropriate powers of the
temperature.
Since the temperature is the only scale available the energy density must
have the same $T$-dependence as for an ideal gas
\be
\rho=cT^4
\label{Sintra27}
\ee
Only the coefficient $c$ should be different from the value obtained by
counting the perturbative degrees of freedom\footnote{A similar remark
also applies to high temperature QCD. We expect quantitative modifications
of early cosmology due to the difference between $c$ and the ideal gas
value.}. We expect that quarks and leptons form $SU(2)$ singlet bound
states similar to the mesons in QCD\footnote{We use here a language
appropriate for the excitations of the three-dimensional Euclidean
theory. Interpretation in terms of relativistic particles has to be used
with care!}. A chiral condensate seems, however,
unlikely in the high-temperature regime and we do not think that fermions
play any important role for the dynamics of the electroweak
phase-transition. The ``photon'' (or rather the gauge boson associated to
weak hypercharge) decouples from the $W$-balls. Its effective high
temperature
coupling to fermion and scalar bound states is renormalized to a very
small
value. As for the phase with spontaneous symmetry breaking, the fermions
and ``photon'' can be neglected for the symmetric phase. We conclude
that the high-temperature phase-transition of the electroweak theory can
be described by an effective three-dimensional Yang-Mills-Higgs system.
It is strongly interacting in the symmetric phase. Depending on the value
of the mass of the Higgs boson it may also be strongly interacting  in
the phase with spontaneous symmetry breaking if the temperature is near
the critical temperature. A more detailed investigation of this issue will
be given in the next section.

\bigskip\noindent
{\large\bf{4. Renormalization Group improved Effective Potential for the
Electroweak Phase-Transition}}
\bigskip\\
For an approximate study of the electroweak phase-transition
we will now investigate a three-dimensional $SU(2)$-Higgs model
\cite{ourselves}.
It is related to the full electroweak theory at high temperatures
by setting the $U(1)_Y$-coupling to zero and integrating out
the non-static modes of all fields as well as all modes of the
$0$-component of the gauge-field. For an extensive discussion
of this procedure, see
\cite{J}.
We will work with the truncation
\be
\Gamma_k \left[ \varphi, A_\mu ,\bar{A}_\mu\right] &=&
\int d^dx \left( U_k(\rho) +
Z_{\varphi,k} | D_\mu \varphi |^2 +
\frac{1}{4} Z_{F,k} F_{\mu\nu} F^{\mu\nu} \right.\nonumber \\
& & \left. +\frac{1}{2\alpha} Z_{F,k} \left( D_\mu[\bar{A}](A^\mu-\bar{A}^\mu)
\right)^2
\right)
\label{alt_1}
\ee
and a simple truncation in the ghost sector
\cite{Sintra_15}.
Here $\rho = \varphi^\dagger
\varphi$ and group indices are omitted.
The form of the average potential
$U_k(\rho)$ is left arbitrary
and has to be determined by
solving the flow equation.
For the Abelian Higgs model the
evolution equation for the
average potential was computed
in the approximation (\ref{alt_1})
in reference
\cite{Sintra_15}.
Inserting the appropriate
$SU(2)$ group factors
we obtain in the Landau
gauge ($\alpha=0$), with
$t=\ln k$
\be
\frac{\partial}{\partial t} U_k(\rho) &=&
\frac{1}{2} \int\!\!
\frac{d^3 q}{\left( 2 \pi \right)^3}
\frac{\partial}{\partial t}
\left( 6
\ln \left( q^2+k^2+ m_B^2 \right) + \right. \nonumber \\
& & \left. + \ln \left( q^2+k^2 + m_1^2 \right) +
3 \ln \left( q^2+k^2 + m_2^2 \right) \right)
\label{alt_2}
\ee
where the mass terms read
\be
m_B^2 = \frac{1}{2} Z_{\varphi,k} \bar{g}_3^2 \rho \ ; \
m_1^2 = \left( U_k'(\rho) +
2 \rho U_k''(\rho) \right) / Z_{\varphi,k} \ ; \
m_2^2 = U_k'(\rho) / Z_{\varphi,k} .
\label{alt_3}
\ee
We have used here a masslike
infrared cutoff $R_k = Z_k k^2$.
The partial derivative
$\frac{\partial}{\partial t}$ on the
right hand side of (\ref{alt_2}) is
meant to act only on $R_k$ and
we omit contributions arising
from the wave function renormalization
$Z_k$ in $R_k$. Primes denote
derivatives with respect to $\rho$.

This flow equation\footnote{
The ultraviolet divergence on the
right hand side of equation (\ref{alt_2}) is
particular to the use of a masslike
infrared cutoff. For $d = 3$ it
concerns only an irrelevant constant
in $U$ and is absent for
$\partial U'/\partial t$.}
constitutes
a nonlinear partial differential
equation for the dependence
of $U$ on the two variables
$k$ and $\rho$.
In our case it holds for the
three dimensional potential
$U_3$ and a correspondingly
normalized scalar field
$\rho_3$. They are related to
the usual four dimensional
quantities by $U_3 = U_4 / T$ ,
$\rho_3 = \rho_4 / T$.
Equation (\ref{alt_2}) is the basic
equation of this section and has to
be supplemented by corresponding
equations for the $k$-dependence
of $\bar{g}_3^2$ and $Z_\varphi$.
For given initial conditions at
$k_T$ we aim for a solution for
$k\rightarrow0$ in order to
compute the free energy $U_0$.
The gauge coupling in (\ref{alt_3})
stands for the three dimensional
running renormalized gauge coupling
$\bar{g}_3^2(k)$. Its value
at the scale $k_T$ is given by
\be
\bar{g}_3^2(k_T) = g_4^2(k_T) T
\left( 1 - \frac{g_4^2(k_T) T}
{24 \pi m_D} \right)
\label{alt_4}
\ee
where
\be
m_D^2 = \frac{5}{6} g_4^2(k_T) T^2
\label{alt_neu1}
\ee
accounts for the effects of integrating
out the $A_0$ mode
in lowest order
\cite{J}.
The evolution equation for the
running gauge coupling in the pure
Yang-Mills theory has been given
above (eq. (\ref{Sintra21})), and
reads in lowest order
\be
\frac{\partial}{\partial t} \gdzr = \beta_{g^2} =
- \frac{23 \tau}{24 \pi} {\bar{g}_{3}}^4(k) k^{-1} \,\, .
\label{alt_5}
\ee
The deviation of $\tau$
from one accounts for the
small
contributions of scalar fluctuations
which are not included here\footnote{
For a suitable choice of wave function
renormalization constants in the infrared
cutoff for the gauge bosons the lowest
order result becomes independent of
the gauge parameter $\alpha$ and can therefore be
used for the Landau gauge employed
here.}.
Equation (\ref{alt_5}) is easily solved,
\be
\frac{1}{\gdzrk} = \frac{1}{\gdzrkt} +
\frac{23 \tau}{24 \pi} \left( \frac{1}{k_T} -
\frac{1}{k} \right)
\label{alt_17}
\ee
and yields the confinement scale
(cf. eq.(\ref{Sintra26})) in lowest order
\be
\Lambda^{(3)}_{\rm conf} = \left( \frac{1}{k_T} +
\frac{24 \pi}{23 \tau \gdzrkt}
\right)^{-1} \,\,.
\label{alt_18}
\ee
Furthermore, we will need the anomalous dimension
of the scalar field. For
our purpose it can be approximated by
\cite{Sintra_15}
\be
\eta_\varphi = - \frac{\partial \ln Z_\varphi}{\partial t}
 = - \frac{1}{4\pi} \gdzrk k^{-1}
\label{alt_6}
\ee
and we set $Z_\varphi=1$ for $k=k_T$.

A convenient quantity for an investigation
of the effective potential is the $\rho$-dependent
quartic coupling
\be
\bar{\lambda}_{3,k}(\rho)
= U_k''(\rho)
= \frac{\partial^2 U_{3,k}}{\partial \rd^2} .
\label{alt_7}
\ee
Knowing for $k=0$ the function
$\bar{\lambda}_3(\rho) =
\bar{\lambda}_{3,0}(\rho)$,
the high temperature effective
potential $U(\rho) = U_0(\rho)$
can be reconstructed by integration
and translation to a four
dimensional normalization.
One of the two integration
constants is irrelevant and
the other (the mass term
linear in $\rho$) can be found by adapting
$U(\rho)$ to the perturbative
result for large $\rho$ where
the three-dimensional
running of the couplings is irrelevant.

The evolution equation for the $k$-dependence
of $\bar{\lambda}_3(\rho_3)$ can
be inferred from (\ref{alt_2}) by
differentiating twice with respect to
$\rho_3$ and reads
\be
\frac{\partial \bar{\lambda}_{3,k}
(\rd)}{\partial t}  = \frac{3}{32 \pi} \!\left(
\frac{Z_{\varphi,k}^2 \bar{g}_{3,k}^4 k^2}
{\left( k^2+m_B^2 \right)^{3/2}} \!+\!
\frac{6 Z_{\varphi,k}^{-2} \bar{\lambda}_{3,k}^2(\rd) k^2}
{\left( k^2+m_1^2 \right)^{3/2}} \!+\!
\frac{2 Z_{\varphi,k}^{-2} \bar{\lambda}_{3,k}^2(\rd) k^2}
{\left( k^2+m_2^2 \right)^{3/2}} \right)
\label{alt_10}
\ee
where we have neglected terms
$\propto U_k^{(3)}(\rd)$ and
$U_k^{(4)}(\rd)$.
We report here on
an approximate solution of the flow equation
for $\bar{\lambda}_3(\rd)$ for $k=0$
\cite{ourselves}.
It is based on the observation that
the \rd-dependent mass terms
$m_B^2$, $m_1^2$, and $m_2^2$ act in
equation (\ref{alt_2}) as independent infrared
cutoffs in just the same way as $k^2$.
A variation of $m^2$ for $k^2=0$ is
roughly equivalent to a variation
of $k^2$ at $m^2=0$.
We use this observation to translate the flow equation
(\ref{alt_10}) into a renormalization
group equation for $\ld(\rd)$ at
$k=0$: In equation (\ref{alt_10}) we
replace $\frac{\partial}{\partial t}$ by
$\frac{\partial}{\partial t'} =
m_B \frac{\partial}{\partial m_B}$ and the factors
$k^2 \left(k^2+m^2 \right)^{-3/2}$ by
$m^{-1}$.
We can then work with a
new effective infrared cutoff $k'=m_B$
which is a function of \rd\
(we omit the prime on $k$ in the
following),
\be
k^2 = m_B^2 = \frac{1}{2} Z_\varphi(k) \gdzk \rd .
\label{alt_13}
\ee
This procedure transforms equation (\ref{alt_10})
into a simple differential equation
for $\bar{\lambda}_3(\rd) =
\bar{\lambda}_3(k(\rd))$. In terms
of the renormalized coupling
\be
\ldrk = Z_\varphi^{-2}(k) \ldk
\label{alt_14}
\ee
it reads\footnote{
A more formal justification for eq.~(\ref{alt_15})
can be found in
\cite{ourselves}.}
\be
\frac{\partial}{\partial t} \ldrk =
\frac{3}{32 \pi k} \left(
\bar{g}_3^4(k) +
\left( \sqrt{6} + \sqrt{2} \right) \bar{\lambda}_R^{3/2}(k)
\bar{g}_3(k) \right)
- \frac{1}{2 \pi k} \gdzrk \ldrk .
\label{alt_15}
\ee
For the terms $\propto \frac{1}{m_1}$
and $\propto \frac{1}{m_2}$ from
equation (\ref{alt_10})
we have approximated
in (\ref{alt_3}) $U'(\rd) \simeq
\rd \bar{\lambda}_R(\rd)$ which amounts
to neglecting the mass term. For negative
\ldrk\ our approximation does
not describe properly the effect of the
scalar fluctuations. Since their
contribution is small in this
region we simply omit the terms
$\propto \bar{\lambda}_R^{3/2}$
once \ldrk\ becomes negative.

In order to solve the flow equation (\ref{alt_15})
we need an initial value $\bar{\lambda}_3(k_T)$,
in addition to (\ref{alt_4}).
This will depend on the (zero temperature) scalar mass $M_H$.
Furthermore, the $\rho$-integration of
(\ref{alt_7}) leads to a term $-\mu^2(T) \rho_3$
with the temperature dependent mass term $\mu^2(T)$ as
an integration constant. The values of $\bar{\lambda}_3(k_T)$ and
$\mu^2(T)$ describe how a given four-dimensional
model at $T>0$ translates into an equivalent three-dimensional
one. They can be fixed by the observation that for large
$\rho$, such that $m_B^2(\rho)>k_T^2$, the one-loop
expression for the effective potential
\cite{???}
\be
\tilde{U}_3(\rd) = -\mu^2(T) \rd \!+\!
\frac{1}{2} \!\left( \ld + \Delta \ld
\right) \!\rd^2 \!-\! \frac{1}{12 \pi} \!\left(
6 m_B^3 \!+\! 3 m_E^3 \!+\! \bar{m}_1^3 \!+\! 3 \bar{m}_2^3 \right)
\label{old26}
\ee
should be a good approximation. Here the masses are given by
$\bar{m}_1^2 = 3 \ld \rd - \mu^2(T)$,
and $\bar{m}_2^2 =  \ld \rd - \mu^2(T)$, and we set
$Z_\varphi = 1$.
Also the correction
\be
\Delta \ld = \frac{3 \bar{g}_3^4}{
64 \pi^2 T} \left( 1 +
\frac{\sqrt{6}+\sqrt{2}}{8}
\frac{M_H^3}{M_H^3} \right)
\label{old27}
\ee
is chosen such that $\ld = \tilde{U}_3''(8 \pi^2 T^2 / \gdzkt)$.
High temperature perturbation theory to one loop with the
$A_0$ integrated out leads then to the initial value
\be
\bar{\lambda}_3(k_T) =
\frac{1}{4} g_4^2(k_T) T
\frac{M_H^2}{M_W^2} -
\frac{3 g_4^4(k_T) T^2}{64 \pi m_D} -
\Delta \ld \,\, .
\label{old3}
\ee
We observe that the inclusion of two-loop effects or
fermions will change the relation between $\bar{\lambda}_3(k_T)$
and $M_H$. This leads to a rescaling between the scalar mass
quoted in this work and the true physical mass. The mass term
$\mu^2(T)$ has in two-loop perturbation theory the genuine
temperature dependence
\be
\mu^2(T) = \beta M_H^2 - \gamma T^2 + \bar{\gamma} T^2 \ln
\frac{T}{M_W} \,\, .
\label{In1}
\ee
Here $\beta$ and $\gamma$ are independent of $T$
\footnote{To one loop order one has ($A_o$ integrated out, no
quarks, $\alpha_w=\frac{g_4^2(k_T)}{4\pi}$)
\be
\beta=\frac{1}{2} \,\, , \,\,
\gamma = \frac{\pi}{4}\alpha_w\left( 3-\frac{3 m_D}{\pi T}+
\frac{M_H^2}{M_W^2} \right) \nonumber
\ee
and no $T^2 \ln T$ term.}.
They depend, however, on the regularization scheme and this is
particularly important for $\gamma$ as indicated
already by a possible change of scale in the $\ln T$-term.
Some care is needed for the proper comparison between three-
and four-dimensional lattice regularizations,
high temperature perturbation theory in various versions and
our renormalization group approach. The most reliable way
of comparison seems to us to equate renormalized quantities in the
various approaches. A good candidate for fixing $\beta$ and $\gamma$
seems to be the expectation value $\Phi_0(T)$ at two different
temperatures $T_1$ and $T_2$ \footnote{
Here $T_{1,2}$ should be sufficiently below the critical
temperature such that the potential minimum occurs
in a region of $\rho$ where two-loop perturbation theory
is reliable.
For purposes of comparison with three-dimensional lattice
results we have chosen $\beta$ and $\gamma$ such as to
obtain the same $\Phi_0(T_{1,2})$ as in two-loop
perturbation theory, with $T_1$ and $T_2$ in a region
where lattice data and perturbation theory are in good agreement
for the prediction of the location of the potential minimum.}.

Combining equations
(\ref{alt_7}),
(\ref{alt_13}), (\ref{alt_14}) and (\ref{alt_15})
with flow equations for $\bar{g}_3^2$ and
$\eta_\varphi$, we can
compute the \rd-dependence of the
high temperature effective potential
as a solution of the flow equation.
It is interesting to note that except
for the last term
arising from the anomalous dimension,
eq. (\ref{alt_15})
can also be directly obtained by taking
appropriate derivatives of the one
loop formula (\ref{old26}),
treating mass ratios such as
$m_B/m_1$ as $k$-independent and
replacing at the end the couplings
\gdz\ and \ld\ by running couplings
evaluated at the scale $k$.
For $Z_\varphi=1$ and \gdz, \ld\ independent
of $k$ equation (\ref{alt_15}) reproduces
exactly the one loop result.
Our renormalization group improvement enables
us to include the effects of running
couplings and the anomalous dimension.
This should reproduce a large part of the two-loop
corrections and also higher contributions.
The main
changes as compared to the 1-loop calculations
can be understood
from the corresponding differential equations:
The inclusion of $\eta_\varphi$ lowers the scale
at which $U''$ changes
sign, whereas the running of the gauge coupling
acts the opposite way. Thus at a given $k$,
corresponding to a given \rd, the running of
\gdzr\ makes the potential bend up less
than the 1-loop calculation predicts.
The inclusion of the term
$\propto \bar{\lambda}_R^{3/2}\bar{g}_3$ will
have the same effect but is quantitatively
important only for a
large scalar mass.

Translated to the effective
potential the running of
\gdz\ should strengthen the phase-transition and lower
the critical temperature as compared to the
one-loop result. This is what one would
naively expect, since the first order character of the transition
is due to the gauge-boson loops.
Thus, enhancing the coupling should give
a transition more strongly first
order. On the other hand,
non-perturbative condensation phenomena may
have the opposite effect. Also the inclusion of
the anomalous
dimension $\eta_\varphi$ weakens the transition. To estimate the
precise effects of the running of the couplings
on the shape of the potential we
proceed to a numerical investigation
of equations (\ref{Sintra21}), (\ref{alt_6}), and
(\ref{alt_15}).
(For all numerical work we use
$g_4 = 2/3$,
$M_w = 80.6 \mbox{\,\,GeV}$, and
$\tau = 1$.)
Additional effects from condensation phenomena
will be qualitatively discussed and added later.
In fig.s\,2, 3 and 4 we show the effective potential as
obtained with our method for different
temperatures and masses of the Higgs scalar.
In all plots we use
four dimensional quantities and plot
$\delta U = U_4(\Phi)-U_4(0)$ versus
\hspace*{1cm}\epsfig{file=Abb2paris.eps,height=11cm}
\vspace*{-0.2cm}\figcaption{The effective potential for $M_H=80$ GeV with
possible contributions from condensation phenomena. All
curves are at their respective critical temperature. The
curve denoted by (1) is the potential without nonperturbative
effects ($a=0$, $T_c=172.8$ GeV). The other curves correspond to
the critical temperature obtained by lattice simulations
\cite{Elweak_lattice_80}, $T_c=167.7$ GeV, and the following
values of the parameters for the nonperturbative effects:\\
\noindent(2) - $a=0.2928$, $b=3$, $c=1.1$, $\alpha=1$ (solid line);\\
\noindent(3) - $a=0.2885$, $b=3$, $c=1$, $\alpha=2$ (long dashed line);\\
\noindent(4) - $a=0.305$, $b=1$, $c=1$, $\alpha=1$ (short dashed line);\\
\noindent(5) - $a=0.3005$, $b=1$, $c=1$, $\alpha=2$ (dotted line).\\
\noindent The lower dashed-dotted line gives the perturbative potential
at the lattice-critical temperature.
\vspace*{0.7cm}}

\noindent $\Phi=\sqrt{\rho_4}$. The first dashed-dotted
curve in figure 2 (denoted by (1)) corresponds to the critical temperature
which would be obtained from our renormalization group-improved
approach, neglecting condensation effects.
The second dashed-dotted curve in figure 2 gives the analogous
result for a temperature corresponding to the critical
temperature inferred from lattice simulations
\cite{Elweak_lattice_80}.
We will now discuss possible alterations due to
nonperturbative condensates and demonstrate how they could
lead to agreement between our method and lattice results.

The effective potential shown in figures 2 through 4 is
expected to be rather reliable for large enough values
of $\Phi$ where the effective gauge coupling is not
yet too strong. On the other hand, the
truncation (\ref{alt_1}) becomes insufficient for large
$g$ and we expect important modifications for $\Phi<\Phi_{\rm np}$,
where (eq.s~(\ref{Sintra26}) and (\ref{alt_13}))
\be
\Phi_{\rm np}&=&\left( \frac{2 k_{\rm np}^2 T}{Z_\varphi(k_{\rm np})
\bar{g}_3^2(k_{\rm np})} \right)^{\frac{1}{2}} \nonumber \\
|\eta_F(k_{\rm np})|&=&1 \,\, .
\label{Z1}
\ee
In fact, the flow equations
(\ref{Sintra21}), (\ref{alt_10}) do not account for
effects like the $W$-condensation mentioned in the
preceding section.
An easy way to visualize the relevance of such effects is the
introduction of a composite $SU(2)$-singlet field $\chi$ for the
description of condensation phenomena in the gauge sector.
For example, one may choose $\chi \propto F_{ij}F^{ij}$ or
some other (properly regularized) operator.
Condensation phenomena are then described by the vacuum expectation
value $\chi_0$. With an appropriate normalization of $\chi$ they
give a contribution to the free energy
\be
\Delta U_3 = \chi_0^3 \,\, , \,\, \Delta U = \chi_0^3 T \,\, .
\label{Z2}
\ee
For $\Phi=0$ the only available scale is $k_{\rm np}\propto T$, and
therefore $\chi_0=a k_{\rm np}$.
The dimensionless coefficient $a$ is expected of order one since the
gauge coupling grows very rapidly for $k<k_{\rm np}$ and there is not much
difference between $k_{\rm np}$ and $\Lambda^{(3)}_{\rm conf}$
(c.f.~eq.~(\ref{Sintra26})).
It is also clear that the value of $\chi_0$ must depend on $\Phi$:
For large $\Phi$ condensation phenomena should be essentially absent since
the gauge coupling remains small. We will parameterize the $\Phi$-dependent
condensation effects by an additional contribution to the free energy
\be
\Delta U = - \left[ a k_{\rm np} f\left(\frac{g^2(k)}{g^2_{\rm np}}\right)
\right]^3 T
\label{Z3}
\ee
where $f$ describes how $\chi_0$ depends on the effective gauge-coupling,
with $f(z\rightarrow \infty)=1$ and $f(z)$ vanishing rapidly for
$z\ll 1$. As an example we take
\be
f(z) = \frac{2}{\pi}{\rm arctan}\left(\frac{\pi}{2}b(z-c)^\alpha\right)
\label{Z4}
\ee
for $z>c$ and $f=0$ otherwise.
Here $c$ indicates for which $g^2$ the condensation sets in, and $b$ is a
measure
how fast the condensation phenomena build up.
The $\Phi$-dependence of $\Delta U$ arises indirectly through
the $\Phi$-dependence of $g$ via the identification (\ref{alt_13}).

Condensation phenomena lower the free energy around the origin
($\Phi=0$) and therefore lead to a lower critical temperature
\cite{Sintra_24,ourselves}.
This is consistent with the fact that the critical temperature
computed without $\Delta U$ comes out systematically higher than found
in lattice simulations. One may even use the lattice results
for $T_c$ to give a rough estimate of the coefficient $a$. Employing
(\ref{Z3}),(\ref{Z4}) with $b=c=1$ we have adapted $a$ such that the
critical temperature coincides with the central values from lattice
simulations. This yields
$a \sim 0.3$ for the simulation of
\cite{Elweak_lattice_80}
at $M_H=80$ GeV.
An attempt for an estimate of the size of
condensation effects for $M_H=35$ GeV yields
$a \sim 0.38$ \cite{Elweak_lattice_35}.
Since the condensation effects are mainly related
to the gauge field degrees of freedom and also the
scalar contribution to the running of $g^2$ is
small one expects $a$ to be independent of $M_H$ in a
first approximation
\cite{ourselves}.
This conjecture seems to be consistent within the large
uncertainties of the quoted values.
\hspace*{1cm}\epsfig{file=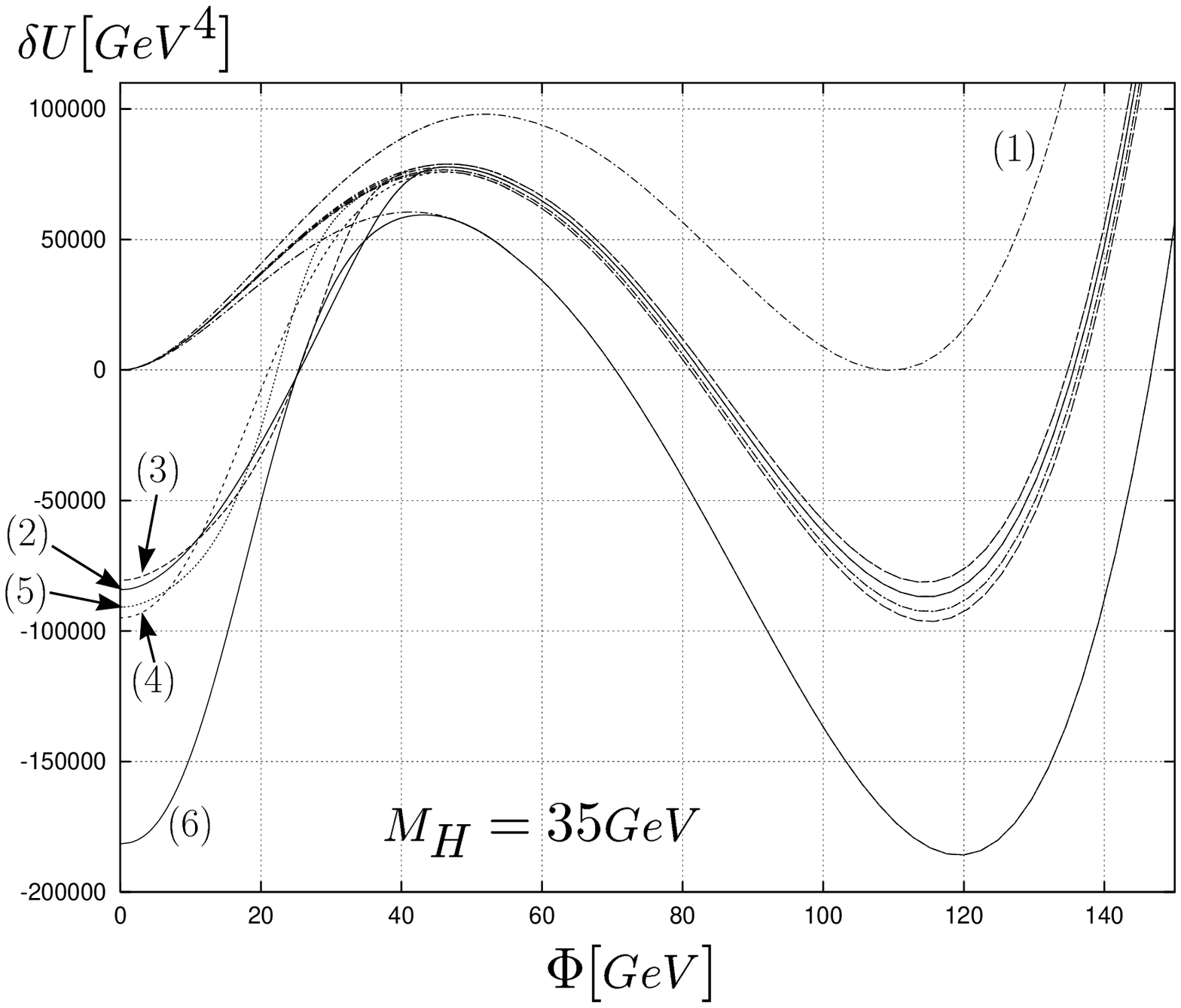,height=11cm}
\vspace*{-0.2cm}\figcaption{As figure 2, but with $M_H=35$ GeV.\\
\noindent(1) - $a=0$, $T=94.61$ GeV (dashed-dotted line);\\
\noindent(2) - $a=0.2928$, $b=3$, $c=1.1$, $\alpha=1$, $T=94.13$ GeV (solid
line);\\
\noindent(3) - $a=0.2885$, $b=3$, $c=1$, $\alpha=2$, $T=94.16$ GeV (long dashed
line);\\
\noindent(4) - $a=0.305$, $b=1$, $c=1$, $\alpha=1$, $T=94.08$ GeV (short dashed
line).\\
\noindent(5) - $a=0.3005$, $b=1$, $c=1$, $\alpha=2$, $T=94.10$ GeV (dotted
line).\\
\noindent The curve denoted by (6) corresponds to the value of $a$
as obtained from lattice simulations for $M_H=35$ GeV
\cite{Elweak_lattice_35}: \\
\noindent(6) - $a=0.381$, $b=1$, $c=1$, $\alpha=1$, $T=93.62$ GeV (solid line).
\vspace*{0.7cm}}

For $\Phi=0$
we note that $\chi_0$ roughly equals the three-dimensional
confinement scale for $a\sim 0.3-0.4$.
Comparison with the curves for $a=0$ (dashed line in figure 2)
demonstrates the importance of the nonperturbative phenomena
for $M_H=80$ GeV.
Our lack of quantitative knowledge of the condensation
phenomena is reflected by the difference of the curves
for $\alpha=1$ and $2$ and $b=1$ and $3$.
For $\alpha=2$ (curves (3) and (5) in figure 2), $\Delta U$ gives
no contribution to the massterm at the origin and we
observe the minimum at $\Phi>0$ even for the ``symmetric'' phase.
This would lead to an effective magnetic mass as described
earlier in the context of gap-equations
\cite{Elweak_gap}.

For $M_H=80$ GeV we observe that condensation phenomena
may weaken or strengthen the phase transition, depending
on the values of $\alpha$ and $b$. As a general tendency
we observe that for a fast onset of the condensation
(large $b$) the phase transition becomes weaker than one
would expect from perturbation theory (compare curves (1) and (2)).
In fig.\,3 we give for comparison the potential at the
critical temperature for $M_H=35$ GeV, with the same
parameters $a$, $b$, $c$, and $\alpha$ as in figure 2.
In addition, we also present a curve for $a=0.381$ (denoted as (6) in the
figure)
\cite{Elweak_lattice_35}. For such a low value of the scalar mass
one expects perturbation theory to be rather
\hspace*{1.2cm}\epsfig{file=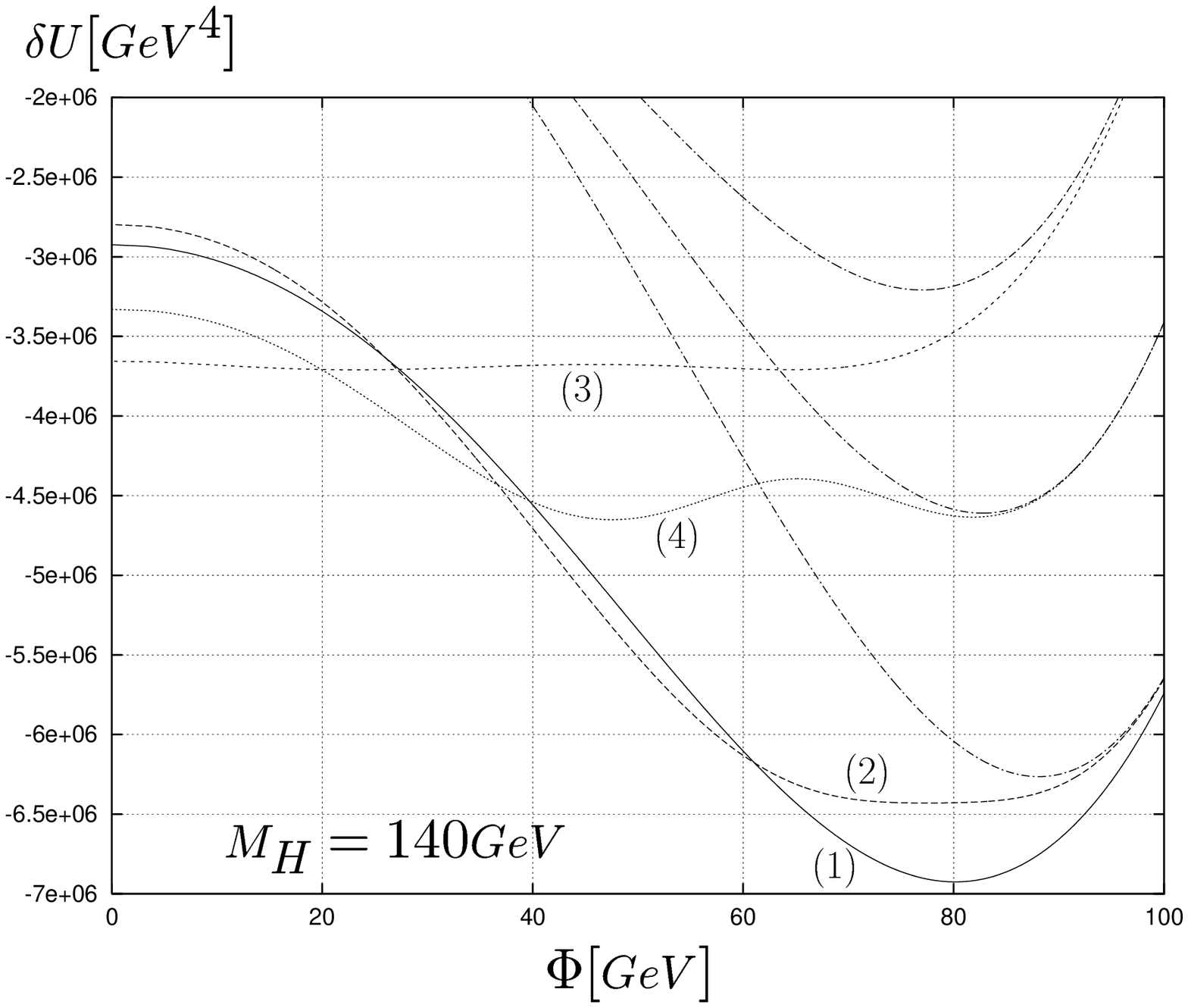,height=11cm}
\figcaption{As figure 2, but for $M_H=140$ GeV. Here only the
curves corresponding to the critical temperatures with
nonvanishing condensation effects are
given. The dashed-dotted lines give the potential at the respective
temperatures neglecting condensation phenomena. The parameters are:\\
\noindent(1) - $a=0.2928$, $b=3$, $c=1.1$, $\alpha=1$, $T=228.5$ GeV (solid
line);\\
\noindent(2) - $a=0.2885$, $b=3$, $c=1$, $\alpha=2$, $T=228.5$ GeV (long dashed
line);\\
\noindent(3) - $a=0.305$, $b=1$, $c=1$, $\alpha=1$, $T=234.3$ GeV (short dashed
line).\\
\noindent(4) - $a=0.3005$, $b=1$, $c=1$, $\alpha=2$, $T=231.5$ GeV (dotted
line).\\
\noindent For the choices of parameters corresponding to (1) and (2), the
transition has changed to an analytical crossover.}

\noindent reliable.
This is confirmed by the fact that the critical temperature
$T_c$ and the location of the minimum $\Phi_0(T_c)$
depend only very moderately on the condensation effects.
Only the barrier height, related to the surface tension,
depends substantially on $a$, being nevertheless almost
independent of the precise form of $f$ as encoded in
$b$, $c$, and $\alpha$.
For a fixed temperature the condensation effects influence the
shape of the potential only for $\Phi \klgl 40$ GeV, far
away from the location of the minimum. The strength of the transition
increases with $a$, independent of the shape of $f$.

Finally, we also try an extrapolation to larger values of
the scalar mass, as shown in figure 4 for $M_H=140$ GeV.
At the critical temperature the shape of the potential
now depends very strongly on the shape of $f$ and an
understanding of the condensation phenomena becomes
crucial even for the qualitative picture. We observe that
all curves for $a\sim 0.3$ have for the symmetric phase a minimum at
$\Phi_0>0$. As $b$ is increased, this minimum moves toward
the minimum corresponding to the phase with spontaneous
symmetry breaking. For the curve with $\alpha=1$, $b=3$
the two minima have already melted and there remains
no true phase transition (solid line). For this form of $f$ one would
predict an analytical crossover
for $M_H=140$ GeV.
This illustrates the speculation
\cite{Sintra_15}
that, as a function of $M_H$, the critical line
corresponding to the first order transition ends at some
value $M_{H,c}$. For this critical value of the
scalar mass the phase transition would have to be of
second order\footnote{At this transition one would
expect interesting behavior determined by nontrivial
critical exponents. This is beyond our approximations.},
with vanishing scalar mass at the critical temperature.
By accident, this situation is realized approximately for
$\alpha=2$, $b=3$ (i.e.\,\,$M_{H,c} \sim 140$ GeV in this case).
It would be very interesting to know the true value of
the critical Higgs-mass!

Comparing figures 2, 3, and 4 we have learned that the
importance of non-perturbative condensation effects
strongly increases with $M_H$.
We conclude that perturbation theory can only give
a realistic description of the phase transition for a
small scalar mass, whereas the nonperturbative effects are
of crucial importance for the understanding of the
electroweak phase transition for realistic values of
the scalar mass.\bigskip\\

\end{document}